\documentclass[prl,twocolumn,superscriptaddress,amsmath,showpacs]{revtex4} % ,showpacs,superscriptaddress,preprintnumbers,amsmath,amssymb

\usepackage{graphicx}% Include figure files
\usepackage{dcolumn}% Align table columns on decimal point
\usepackage{bm}% bold math\begin{document}

\begin{document}

%\preprint{APS/123-QED}

\title{Universal mesoscopic statistics and the localization of light}% Force line breaks with \\

\author{Jongchul Park}%
\affiliation{Department of Physics, Queens College, The City University of New York, Flushing, NY 11365, USA}%
\author{Sheng Zhang}
\affiliation{Department of Physics, Queens College, The City University of New York, Flushing, NY 11365, USA}
\affiliation{ChiralPhotonics Inc., Pine Brook, NJ 07058, USA}
\author{Azriel Z. Genack}%
\affiliation{Department of Physics, Queens College, The City University of New York, Flushing, NY 11365, USA}%

\date{\today}

\begin{abstract}
We follow the evolution with sample thickness, of intensity statistics for localized light transmitted through layered media in a crossover from one to three dimensions occasioned by transverse disorder. The probability distribution of intensity changes from one dimensional to a mixture of a mesoscopic function of a single parameter, the ``statistical conductance,'' and a distribution of intensity for Gaussian waves. This suggests that the change to a universal statistics beyond 1D is associated with the topological change in the spatial field distribution. 
\end{abstract}


\maketitle

The flow of energy in random systems depends upon the closeness to the threshold for Anderson localization separating diffusing and localized waves. \cite{1,2} On one side of this divide, average transport may be described by the diffusion of particles, such as electrons or photons. On the other side, propagation is strongly suppressed and proceeds via the excitation of spatially localized resonances of the underlying quantum mechanical or classical waves. \cite{3,4,5} The nature of transport in an ensemble of nondissipative random samples can be gleaned from the scaling of the ensemble average of the dimensionless conductance, $g = G/(e^2/h)$, which depends only upon the average value of g and the dimensionality of the scaling. \cite{2} Following convention, $g$ and its ensemble average will be denoted by the same symbol. Localization is achieved for $g\le 1$. The Landauer relation shows that $g$ is equivalent to the optical transmittance, $g = T = \sum_{ab}T_{ab}$, where $T_{ab}$ is the transmission coefficient for an incident transverse mode $a$ into the outgoing mode or a point on the output surface, $b$. \cite{6} Thus localization may be achieved in principle for light as well for electrons. Localizing classical waves is of particular interest because it is possible to study both local and global fluctuations of transmission in random ensembles and to study Anderson localization without the complication of electron interactions. 

Photon localization has been demonstrated in one \cite{7} and two dimensions, \cite{8,a} as well as in quasi-1D samples, \cite{9} in which the length of a locally three dimensional sample with reflecting side walls greatly exceeds the diameter. However, it is difficult to achieve and demonstrate definitively in random three dimensional samples without structural correlation. \cite{10} In contrast to electrons, which exhibit $s$-wave scattering, the leading order of scattering for light is $p$-wave so that the scattering cross section is at a maximum when the scale of the scattering elements is of the order of the wavelength of light, $D\approx \lambda$. This limits the density of scatterers and consequently the minimum achievable transport mean free path for photons, $\ell$. As a result, scattering may not be strong enough to satisfy the Ioffe-Regel criterion for localization in three dimensions, $k\ell = 1$, where $k=2\pi/\lambda$ is the wave vector. \cite{11} At the same time, the exponential scaling of optical transmission \cite{12} is not an unambiguous sign of localization since absorption in diffusive samples also leads to the exponential falloff of transmission. \cite{13}

Because of large fluctuations in mesoscopic systems, which increase dramatically in the localization transition, it is essential to measure the statistics of transport. The probability distributions of total transmission and intensity normalized to their respective ensemble averages, $P(s_a=T_a/\langle T_a\rangle)$ and $P(s_{ab}=T_{ab}/\langle T_{ab} \rangle)$, are directly related in quasi-1D systems because the transmitted wave is thoroughly mixed so that intensity statistics are uniform across the sample output. These distribution were calculated in the diffusive limit for nondissipative samples and expressed in terms of a single parameter, $g$. \cite{14,15} Since $g = 2/3{\rm var}(s_a)$ in this limit, the distribution of $P(s_a)$ depended only upon the variance of the distribution. The functional forms of $P(s_a)$ and $P(s_{ab})$ were found to be valid even in strongly scattering and dissipative quasi-1D samples once the ``statistical conductance'', $g^\prime = 2/3{\rm var}(s_a)$ is substituted for $g$. \cite{16} $g^\prime$ scaled inversely with $L$ for $g^\prime>1$ and fell exponentially for $g^\prime<1$, \cite{9} as was originally predicted for the localization parameter $g$. \cite{2}

Recently, the intensity distribution was compared to measurements of pulsed and steady ultrasound in a slab of randomly positioned aluminum beads sintered to form an elastic network. \cite{17} Measurements of $P(s_{ab})$ were well fit by the distribution in Eq. \eqref{grp} below at different frequencies with a minimum in $g^\prime$ 0.8. Agreement with the distribution found in quasi-1D samples is intriguing in this 3D system since the probability distribution of integrated transmission depends crucially on the area over which intensity is averaged, but there is no specific area over which $s_a$ or $g^\prime$ can be unambiguously defined in the slab geometry.

Since dimensionality is crucial to scaling, \cite{2} a study of transmission statistics in a continuous change of dimension may elucidate the general character of statistics in random media. A change of dimensionality can be induced in random layered media with transverse disorder. \cite{18} Measurements of average transmission in stacks of glass cover slips showed a change from one-dimensional scaling, with an asymptotic exponential scaling for $\langle T(L)\rangle \sim \exp(-L/2\xi)$, where $\xi$ is the average localization length of intensity, to an inverse scaling of transmission, characteristic of diffusion. Disordered layered systems are ubiquitous and present opportunities and challenges in diverse settings such as multilayered Bragg gratings, acoustic scattering in the earth’s crust, and anomalous conductivity anisotropies \cite{20}.  

In this Letter, we follow the evolution of $P(s_{ab})$ for plane wave excitation for localized light in a crossover from one to three dimensions in layered media induced by transverse disorder. In thin samples in which light does not spread beyond a single coherence area, $P(s_{ab})$ is in accord with 1D simulations, and falls sharply to zero as $s_{ab} \rightarrow 0$ and for $T_{ab} > 1$. In thicker samples, the spatial intensity pattern undergoes a topological transformation in which phase singularities are formed at intensity nulls as the lateral spread of the wave exceeds the field correlation length. $P(s_{ab})$ changes to the distribution given in Eq. \eqref{a} below, which was originally found for quasi-1D system. This distribution can be written formally as a mixture of a mesoscopic distribution of $s_a$, $P(s_a)$, and a distribution of intensity for Gaussian waves, even though there is no area for which the distribution of integrated intensity is given by the quasi-1D expression for $P(s_a)$. These measurements show that mesoscopic intensity statistics have a universal form beyond 1D and provide a definitive test for Anderson localization.

The sample is composed of glass cover slips with intervening air gaps. The cover slips are $22\times 22$ $mm^2$ with average thickness of 150 $\mu m$ and standard deviation of the wedge angle of $\sim 0.05^\circ$. The refractive index of glass is 1.522 at the 632.8 $nm$ wavelength of the polarized single-frequency helium-neon laser used in the experiment. The first surface of the sample is at the waist of the normally-incident 5-mm-diameter beam. A tapered optical fiber with mode field diameter of 4 $\mu m$ was centered in the beam and placed just behind the stack. The signal was detected by a PMT. The sample was translated in the layer plane over a $2\times 2-mm^2$ area on a $10-\mu m$ grid to yield a speckle pattern equivalent to that for an incident plane wave. For each sample thickness, speckle patterns were obtained for 10 different arrangements of cover slips. Examples of speckle for 20 and 60 cover slips are shown in Fig. 1.

\begin{figure}
\includegraphics[scale=0.7]{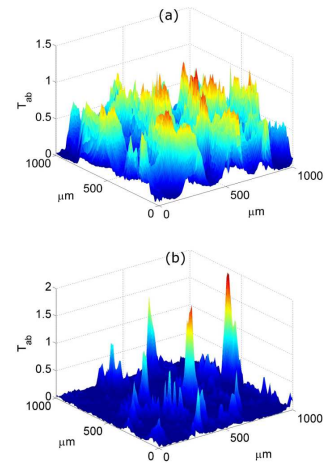}
\caption{\label{Fig1} (Color Online) Typical speckle patterns measured at the output of the samples with (a) $L = 20$ and (b) $L = 60$.}
\end{figure}

Measurements of ${\rm var}(s_{ab})$ versus number of cover slips $L$ are shown in Fig. 2 and compared to 1D transfer matrix simulations. The measured variance is in excellent accord with 1D simulations in thin samples, but departures can be seen for $L>30$. Beyond this thickness, the average transverse spread of the wave, $\sigma_{\perp}$, exceeds the field correlation length, $1/\delta k_\perp$, where $\delta k_\perp$ is the width of the distribution of  $k_\perp$. \cite{18} Thus, for $\delta k_\perp\sigma_{\perp}>1$, wave propagation is no longer one-dimensional and transmission is not determined solely by the longitudinal variation of the index of refraction along a line through the sample but also reflects lateral scattering due to transverse inhomogeneity. The measured values of ${\rm var}(s_{ab})$ reach a maximum value of 3.25 at $L \sim 80$ while the simulated results increase exponentially. This may be compared to the value at the localization threshold in quasi-1D determined by the condition, $g^\prime=1$, ${\rm var}(s_{ab}) = 2{\rm var}(s_a) +1 = 4/3g^\prime + 1 = 7/3$. \cite{14,16}. The question arises as to whether the value $g^\prime = 0.63$ inferred from the above relation with ${\rm var}(s_{ab}) = 3.25$ signifies that the wave is localized in the slab geometry of a layered medium. This question may be addressed by examining the full intensity distribution as $L$ is changed and comparing the results to the quasi-1D distribution.

\begin{figure}
\includegraphics[scale=0.7]{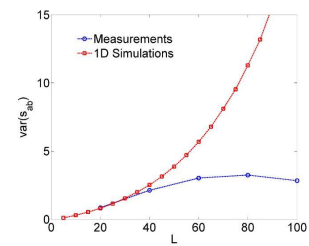}
\caption{\label{Fig2} (Color Online) Comparison of measured ${\rm var}(s_{ab})$ changing with $L$ to 1D-simulation.}
\end{figure}

The relationship between the statistics of $s_{ab}$ and $s_a$ in quasi-1D follows from the statistical independence of $s^\prime_{ab}\equiv s_{ab}/s_a=$ and $s_a$ and the Gaussian field statistics in single configurations. \cite{14,21} Fluctuations of intensity in quasi-1D samples may therefore be represented as a mixture of $P(s_a)$ and $P(s^\prime_{ab})$. Since $P(s^\prime_{ab})$ is a negative exponential distribution for polarized transmission, $P(s^\prime_{ab})=\exp(-s^\prime_{ab})$, the intensity distribution is given by,
\begin{subequations}\label{grp}
\begin{align}
  P(s_{ab}) &= \int_0^{\infty} \frac{1}{s_a}\exp\left( -\frac{s_{ab}}{s_a} \right) P(s_a)ds_a, \label{a}\\
  P(s_{a}) &= \frac{1}{2\pi i}\int_{-i\infty}^{i\infty} \exp(qs_a)F(q/g^\prime)dq, \label{b}\\
  F(q) &= \exp\left[-g^\prime \ln^2\left( \sqrt{1+q} + \sqrt{q} \right) \right]. \label{c}
\end{align}
\end{subequations}

The distribution $P(s_a)$ is found from RMT \cite{14} and diagrammatic \cite{15} calculations in the weak scattering limit in the absence of dissipation to be a function of a single parameter $g$. The distribution has been found to apply more broadly in quasi-1D to dissipative and strongly scattering systems once $g^\prime = 2/3{\rm var}(s_{ab})$ is substituted for $g$ in the expression for $P(s_a)$, \cite{16}

Measurements of $P(s_{ab})$ in layered samples are shown for $L =$ 20, 30, 60 and 100 in Fig. 3 and compared to 1D simulations and to Eq. \eqref{a}. $P(s_{ab})$ for $L = 20$ is in excellent agreement with 1D simulations up to $s_{ab} = 2.5$. The maximum value of $s_{ab}$, above which 1D simulations of $P(s_{ab})$ vanishes, which is 3.4 for $L = 20$, corresponds to full transmission, $T_{ab} = 1$. Higher values of transmission occur, however, because of the transverse spread of the beam due to nonuniformity in the layers, which does not exist in 1D. $P(s_{ab})$ is expected to be high for small values of $s_{ab}$ for localized waves in 1D samples since transmission is low when the light is off resonances with localized modes and when modes are peaked away from the center of the sample. However, as seen in the inset in Fig. 3(a), $P(s_{ab})$ drops precipitously towards zero as $s_{ab}\rightarrow 0$. The absence of nulls in transmission is a distinctive feature of 1D media and follows from the finite transmission coefficient for flux through the $i/i+1$ interface between parallel layers, $2n_in_{i+1}/(n_i+n_{i+1})^2$ . This requires that the component of flux propagating towards the sample output does not vanish in a layer if it is nonzero in the preceding layer. Nulls in intensity of transmitted light can occur, however, in layered samples once the distribution of transverse wave vectors, $k_\perp$, is broadened by scattering from layers with nonparallel surfaces. 

\begin{figure}
\includegraphics[scale=1]{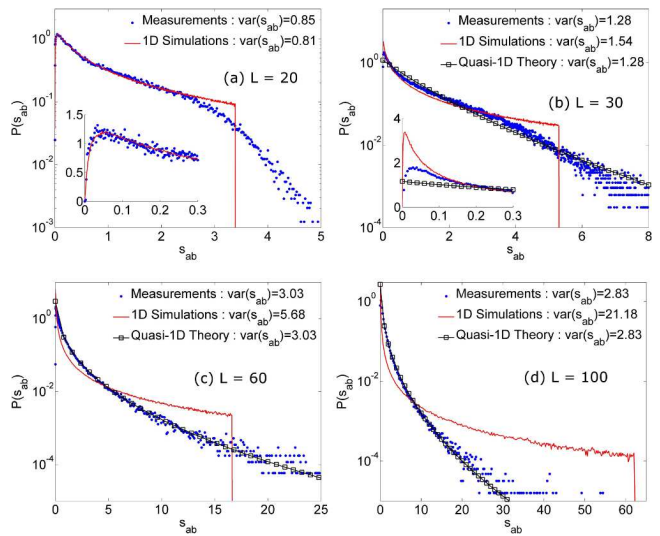}
\caption{\label{Fig3} (Color Online) Comparison of the measured $P(s_{ab})$ to 1D-simulation and Eq. \eqref{a} with same $\rm var(s_{ab})$. Samples studied are with $L =$ 20 for (a), 30 for (b), 60 for (c) and 100 for (d). Inset of (a) shows the details of $P(s_{ab})$ for small $s_{ab}$, demonstrating the absence of phase singularities at $L = 20$.}
\end{figure}

For $L = 30$, the intensity distribution falls between the 1D and quasi-1D distribution, whereas $P(s_{ab})$ for $L = 60$ and 100, are in agreement with Eq. \eqref{a} using the value of $g^\prime$ obtained from the relation, $g^\prime = [{\rm var}(s_{ab})-1]/2$, within the limits set by the mode field diameter of the tapered optical pickup fiber. The agreement with Eq. \eqref{a} in this highly anisotropic medium in a three-dimensional slab geometry is surprising since the statistics of propagation in quasi-1D flow from the complete mixing of waves at the sample output which gives uniform transmission statistics across the output. But, in the slab geometry, the mixture of transmission channels as well as the quasimodes through which the wave is coupled to the output varies along the output. Further, whereas the statistics of $s_a$ depend critically on the sample cross section in quasi-1D samples, there is no well-defined area over which $s_a$ can be defined in the slab geometry. The distributions of integrated transmission through a number of different circular areas at the output of the sample with $L = 60$ are shown in Fig. 4 and compared to the distribution of $P(s_a)$ for the value of $g^\prime$ inferred from the measurements of $P(s_{ab})$. None of the curves for integrated transmission correspond to the distribution predicted by Eq. \eqref{a} for $P(s_a)$. Yet, Eq. \eqref{a} quantitatively describes intensity statistics, once $s_a$ is treated as a variable of integration rather than as a random variable. This suggests that $g^\prime$ is a generalized measure of mesoscopic fluctuations not directly related to the statistics of integrated transmission. It may be related to the probability of crossing of Feynman paths at scattering centers within the medium or the number of electromagnetic quasimodes within the medium contributing to the field at the output. We find that for $L > 40$, $g^\prime$ is equal to the product of the number of coherence areas in an area over which the incident field spreads at the output, $N=(\delta k_\perp\sigma_\perp)^2$ , and the ensemble average transmission coefficient, $g^\prime = N\langle T_a\rangle$ , where, $\langle T_a\rangle =  N  \langle T_{ab}\rangle$. Since, $g = \sum_{ab}T_{ab} = N\langle T_a\rangle$ , in this case, $g^\prime = g$.

\begin{figure}
\includegraphics[scale=0.7]{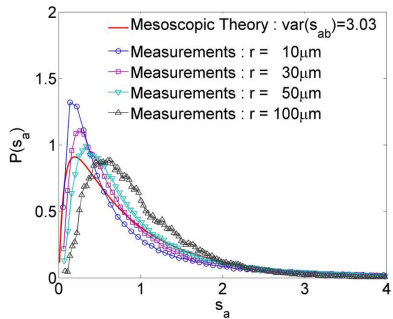}
\caption{\label{Fig4} (Color Online) Probability distributions of intensity integrated over circular areas with different radii for $L=60$. The results are compared with $P(s_{a})$ given by Eq. \eqref{b}.}
\end{figure}

The change of $P(s_{ab})$ with increasing thickness is a consequence of the change of the spatial distribution of intensity. Beyond 1D, $P(s_{ab})$ is obtained by mixing the mesoscopic distribution, $P(s_a)$, with a distribution of intensity for a random Gaussian process [Eq. \eqref{a}]. The change of the spatial distribution of transmitted flux from 1D to higher dimension is a distinctive topological transformation accompanied by the formation of a speckle pattern with intensity nulls which lie at the center of vortices of electromagnetic flux. \cite{22,23,24} Such optical vortices form a network which underpins random Gaussian fields. \cite{25} Low intensity values within the speckle pattern predicted by Eq. \eqref{a} are associated with points within the vortex core. Gaussian statistics arises from the sum of numerous statistically independent contributions to the output field. As the number of transverse modes contributing to the field at a point increases, the assumption of underlying Gaussian statistics for intensity are more nearly satisfied. 

In conclusion, we have examined the statistics of light localized longitudinally by multiple scattering and laterally by variations in the transverse and longitudinal disorder in the sample which create a pattern of transmission resonances. Localization in this anisotropic 3D sample is due to the restricted phase space for k which is a narrow cone around the forward and backward directions. The statistics of light transmitted through layered media with transverse disorder is 1D as long as the transverse spread of the light is smaller than the coherence length. The character of intensity statistics changes, however, once phase singularities appear in the transmitted field. Intensity statistics beyond 1D are then a mixture of the intensity distribution for a Gaussian field and a mesoscopic distribution, which depends only upon the mesoscopic variable $g^\prime$. The similarity of local and global statistics of intensity and evolution of the speckle pattern in quasi-1D \cite{21} suggests that a universal statistics should also exist for aspects of speckle change for any dimension beyond 1D. Thus, the statistics of generalized velocity of phase singularities with frequency shift or internal change of the sample, for example, should be a mixture of the velocity distribution for a Gaussian field and a mesoscopic distribution in terms of a parameter proportional to $g^\prime$. The excellent agreement of Eq. \eqref{grp} with measured intensity distributions even in the case that the speckle pattern is not generic and reflects the underlying disorder, as may be seen in Fig. 1, suggests that Eq. \eqref{grp} is robust and holds for any form of disorder above 1D.

We thank Samuel Gillman for experimental assistance. This work was supported by the NSF under Grant No. DMR-0538350.

\end{document}